\documentclass[useAMS,usenatbib]{mn2e}
\usepackage{graphicx,psfig,cite,epsf}  
\usepackage{times}
\usepackage{subfigure}
\usepackage{textcomp}
\usepackage{multirow}
\usepackage{amsmath}

\def\apjl{ApJ}
\def\apjl{ApJ}                % Astrophysical Journal, Letters
\def\apjs{ApJS}               % Astrophysical Journal, Supplement
\def\ao{Appl.~Opt.}           % Applied Optics
\def\aap{A\&A}                % Astronomy and Astrophysics
\def\src {XTE\,J1810--197}

\title[XTE\,J1810--197 spin-down]{The variable spin-down rate of the transient magnetar XTE\,J1810--197}
\author[Fabio Pintore]{Fabio Pintore$^1$, Federico Bernardini$^{2,12}$, Sandro Mereghetti$^1$, Paolo Esposito$^1$,  \newauthor Roberto Turolla$^{3,11}$,  Nanda Rea$^{4,5}$, Francesco Coti Zelati$^{6,4,10}$, \newauthor Gian Luca Israel$^7$, Andrea Tiengo$^{8,1,9}$, Silvia Zane$^{11}$ \\
$^1$ INAF - IASF Milano, Via E. Bassini 15, I-20133 Milano, Italy \\
$^2$ New York University Abu Dhabi, Saadiyat Island, Abu Dhabi, 129188, United Arab Emirates \\ 
$^3$ Dipartimento di Fisica e Astronomia, Universita di Padova, via F. Marzolo 8, I-35131 Padova, Italy \\
$^4$ Anton Pannekoek Institute for Astronomy, University of Amsterdam, Postbus 94249, NL-1090-GE Amsterdam, the Netherlands \\
$^5$ Instituto de Ciencias de l'Espacio (ICE, CSIC-IEEC), Carrer de Can Magrans, S/N, 08193, Barcelona, Spain\\ 
$^6$ Universita dell'Insubria, via Valleggio 11, I-22100 Como, Italy \\
$^7$ INAF - Osservatorio Astronomico di Roma, via Frascati 33, I-00040 Monteporzio Catone, Roma, Italy \\
$^8$IUSS - Istituto Universitario di Studi Superiori, piazza della Vittoria 15, I-27100 Pavia, Italy\\
$^9$ Istituto Nazionale di Fisica Nucleare, Sezione di Pavia, via A. Bassi 6, 27100 Pavia, Italy \\
$^{10}$ INAF - Osservatorio Astronomico di Brera, Via Bianchi 46, I-23807 Merate (LC), Italy\\
$^{11}$ MSSL, University College London, Holmbury St. Mary, Dorking Surrey, RH5 6NT, UK\\
$^{12}$ INAF $-$ Osservatorio Astronomico di Capodimonte, Salita Moiariello 16, I-80131 Napoli, Italy\\
}
\begin{document}

\maketitle

\begin{abstract}

We have analyzed  {\it XMM-Newton} and {\it Chandra} observations of the transient magnetar \src\, spanning more than 11 years, from the initial phases of the 2003 outburst  to the current quiescent level.
We  investigated the evolution of the pulsar spin period and we found evidence for two distinct regimes:
during the outburst decay, $\dot{\nu}$ was highly variable in the range $-(2-4.5)\times10^{-13}$ Hz s$^{-1}$, while during  quiescence the spin-down rate was more stable at an average value { of $-1\times10^{-13}$ Hz s$^{-1}$}. 
Only during $\sim$3000 days (from MJD 54165 to MJD 56908) in the quiescent stage it was possible to find a phase-connected timing solution, with $\dot{\nu}=-4.9\times10^{-14}$ Hz s$^{-1}$, and a positive second frequency derivative, $\ddot{\nu}=1.8\times10^{-22}$ Hz s$^{-2}$.
These results are in agreement with the behavior expected if the  outburst of \src\ was due to a strong magnetospheric twist.

\end{abstract}

\begin{keywords}
stars: magnetars -- stars: neutron -- X-rays: stars -- magnetic fields  -- pulsars: individual: (XTE J1810--197)
\end{keywords}

\section{Introduction}

{ Magnetars are isolated neutron stars whose persistent   emission and occasional outbursts are powered by magnetic energy (\citealt{duncan92, thompson93,paczynski92}; see also \citealt{mereghetti08,turolla15}).}
\src\ was   discovered with the {\it Rossi X-ray Timing Explorer (RXTE)} as a 5.45 s X-ray pulsar  \citep{ibrahim04} during a bright  outburst in 2003, and associated to a previously known but unclassified {\it ROSAT} source. 
Further multiwavelength observations \citep{woods05short,rea04,halpern08}, led to classify XTE J1810-197 as a magnetar candidate. 

XTE J1810-197 is the prototype of transient members of this class of sources. It likely spent at least 23 years in quiescence (at a flux of $\sim$$7\times10^{-13}$ erg s$^{-1}$ cm$^{-2}${, in the 0.5--10 keV energy band}) before entering in outburst, in the 2003, when the flux increased by a factor of $\sim$100 \citep{gotthelf04}.
For an { estimated} distance of $3.5$ kpc \citep{camilo06,minter08}, the maximum observed luminosity was  $\sim$$10^{35}$ erg s$^{-1}$, but \src\ might have reached an even higher luminosity, since the  initial part of the outburst was missed. \src\ was also the first magnetar from which pulsed radio emission was detected \citep{camilo06,camilo07}.
A large, unsteady spin-down of $\dot{P}\sim10^{-11}$ s s$^{-1}$ was measured during the outburst decay through radio and X-ray observations, which suggested that the surface dipolar magnetic field is $\sim2\times10^{14}$ G  
\citep{gotthelf04,ibrahim04,camilo06}.

The spectrum of \src\ during the outburst has been modeled by several authors with  two or three blackbody components of different temperature. The colder one has been interpreted as the  (persistent) emission from the whole neutron star surface, while the hotter ones have been associated to cooling regions responsible for the outburst \citep{gotthelf04,bernardini09,bernardini11,alford16}. 
The appearance of hot spots could be due to the release of (magnetic) energy deep in the crust, or to Ohmic dissipation of back-flowing currents as they hit the star surface  \citep{perna08,albano10,beloborodov09, pons12}.
The X-ray pulse profile was energy-dependent and time-variable in amplitude, and it could be generally modelled by a single sinusoidal function \citep[e.g.][]{ibrahim04,camilo07,bernardini09,bernardini11,alford16}.

\begin{table}
  \begin{center}
\footnotesize
   \caption{Log of the \textit{{\it XMM-Newton}} and {\it {\it Chandra}}  observations.} 
\scalebox{0.9}{\begin{minipage}{24cm}
      \label{log}
   \begin{tabular}{l c c c  c c }
\hline 
  Obs.    &   Satellite &Obs. ID  & { Epoch$^a$} & Duration\\
No. &            &             &  MJD    & ks  \\
\hline
1 & {\it {\it Chandra}} &4454                                &52878.9386632& 4.3 \\
2 & {\it {\it XMM-Newton}} & 0161360301          &  52890.5595740  &   9.5 \\
 3 & {\it {\it XMM-Newton}} & 0161360401           &  52890.7083079  &   2.1\\
4 & {\it {\it XMM-Newton}} & 0152833201  & 52924.1677914 &  7.0 \\
5 & {\it {\it Chandra}} &5240                             &52944.6289075& 5.4\\
 6 & {\it {\it XMM-Newton}} & 0161360501      &  53075.4952632  &  17.2\\
 7 & {\it {\it XMM-Newton}} & 0164560601      &  53266.4995129  &  26.7\\
 8 & {\it {\it XMM-Newton}} & 0301270501      &  53447.9973027  &  40.0\\
 9 & {\it {\it XMM-Newton}} & 0301270401      &  53633.4453382  &  40.0\\
 10 & {\it {\it XMM-Newton}} & 0301270301        &  53806.7899360  &  41.8\\
 11 & {\it {\it Chandra}} &6660                                &53988.8111877&31.8\\
12 & {\it {\it XMM-Newton}} & 0406800601           &  54002.0627203  &  48.1\\
 13 & {\it {\it XMM-Newton}} & 0406800701        &  54165.7713547  &  60.2\\
 14 & {\it {\it XMM-Newton}} & 0504650201        &  54359.0627456  &  72.7\\
  15 &{\it {\it Chandra}} &7594                                &54543.0034395&31.5\\
16 & {\it {\it XMM-Newton}} & 0552800301         &  54895.5656089  &   4.3\\
 17 & {\it {\it XMM-Newton}} & 0552800201        &  54895.6543341  &  63.6\\
 18 & {\it {\it XMM-Newton}} & 0605990201         &  55079.6256771  &  19.4\\
 19 & {\it {\it XMM-Newton}} & 0605990301         &  55081.5548494  &  17.7\\
 20 & {\it {\it XMM-Newton}} & 0605990401         & 55097.7062563 & 12.0\\
21&  {\it {\it Chandra}} &11102                               &55136.6570779&26.5\\
 22 & {\it {\it Chandra}} &12105                               &55242.6870526&15.1\\
23 & {\it {\it Chandra}} &11103                               &55244.7426533&14.6\\
24 & {\it {\it XMM-Newton}} & 0605990501         &  55295.1863453  &   7.7\\
25 & {\it {\it Chandra}} &12221                               &55354.1368700&11.5\\
 26 & {\it {\it XMM-Newton}} & 0605990601          &  55444.6796630  &   9.1\\
27 & {\it {\it Chandra}} &13149                               &55494.1643981&16.8\\
28 & {\it {\it Chandra}} &13217                               &55600.9885520&16.2\\
 29 & {\it {\it XMM-Newton}} & 0671060101          &  55654.0878884  &  17.4\\
 30 & {\it {\it XMM-Newton}} & 0671060201          &  55813.3872852  &  13.7\\
31 & {\it {\it Chandra}} &13746                               &55976.3735837&22.5\\
32 & {\it {\it Chandra}} &13747                               &56071.3650797&22.1\\
 33 & {\it {\it XMM-Newton}} & 0691070301           &  56176.9826811  &  15.7\\
 34 & {\it {\it XMM-Newton}} & 0691070401           &  56354.1968379  &  15.7\\
 35 & {\it {\it XMM-Newton}} & 0720780201          &  56540.8584298  &  21.2\\
36 & {\it {\it Chandra}} &15870                               &56717.3097928&22.1\\
 37 & {\it {\it XMM-Newton}} & 0720780301            &  56720.9705351  &  22.7\\
38 & {\it {\it Chandra}} &15871                               &56907.9508362&21.7\\
  \hline
\end{tabular}
\end{minipage}}
\end{center}
{ $^a$ Mean time of the observation.}
\end{table}

Here we report on the pulse period evolution  of \src\ exploiting the full set  
of {\it {\it XMM-Newton}} and {\it {\it Chandra}} X-ray observations carried out in the years 2003--2014 during the outburst decay and in the following quiescent period.

\section{Observations and data reduction}
\label{data_reduction}

We made use of 24 {\it {\it XMM-Newton}} and 14 {\it {\it Chandra}} observations of  \src\, totalizing an exposure time of $\sim$$830$ ks (see the log of observations in Table~\ref{log}).

The {\it {\it XMM-Newton}} data were reduced using SAS v. 14.0.0 and the most recent calibrations. We used the data obtained with the EPIC instrument, which consists of one pn camera and two MOS   cameras. For each observation, we selected events with single and double pixel events ({\sc pattern}$\leq$4) for EPIC-pn and single, double, triple and quadruple pixel events for EPIC-MOS ({\sc pattern}$\leq$12). We set `{\sc flag}=0' so to exclude bad pixels and events coming from the CCD edge. The source and background events were extracted from 30$''$ and 60$''$ radius circular regions, respectively. Time intervals with high particle background were removed.

In three observations (7, 13 and 35) we found inconsistent values between the phases of the pulses derived (as described in the next Section) from  the MOS and pn data.  This is due to a known sporadic problem in the timing of EPIC-pn data, causing a shift of $\pm$1 second in the times attributed to the counts. We identified the times at which the problems occurred and corrected the data { by adding (or subtracting) 1 second to the photon time of arrival from the instant when the problem occurred \citep[see][]{martin12}}.

The \textit{{\it Chandra}} observations were reduced using the {\sc ciao} v.4.7 software and adopting the standard procedures. Source events were extracted from a region of 20$''$ radius around the position of  \src\, and background counts from a similar region close to the source.

Photon arrival times of both satellites were converted to the Solar system barycenter  using  the milliarcsec radio position of \src\ (RA = 272.462875 deg,  Dec. = --19.731092 deg,  (J2000); \citealt{helfand07}) and the JPL planetary ephemerides DE405.

\begin{figure}
\center
\subfigure{\includegraphics[width=13.8cm]{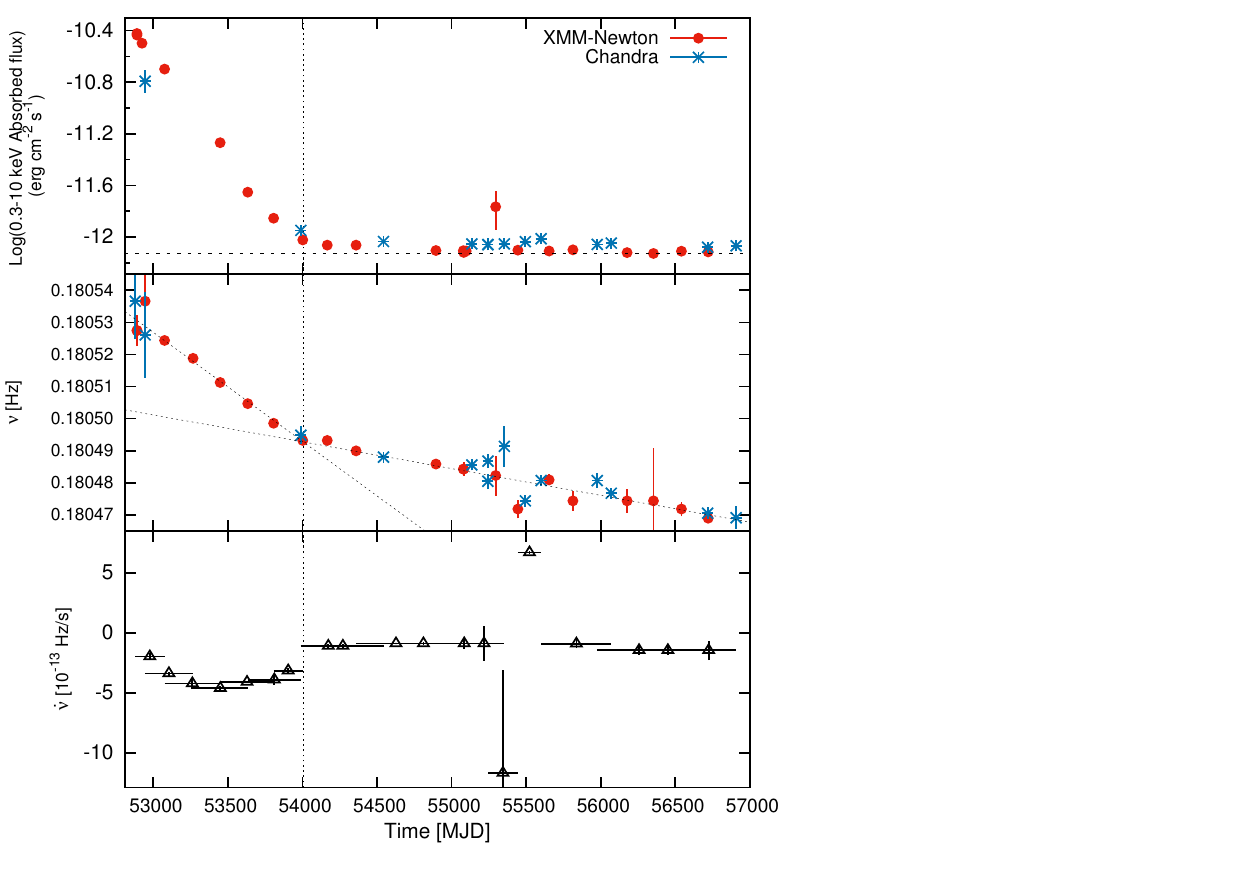}}
\caption{{\it Top panel}: evolution of the logarithmic absorbed flux in the 0.3-10 keV energy range. The dashed line is the linear fit to the data after MJD 54500. The errors are at $90\%$ confidence level. {\it Center panel}: spin frequency along the outburst of \src\ as found in the single observations. The dashed lines indicate the fits with two linear functions to the data before and after MJD 54000. {\it Bottom panels}: frequency derivatives as measured by linear fits of small sub-set of observations. The horizontal error bars indicate the time interval spanned by the observations used in each fit. 
\newline
{ The vertical, dashed line indicates the epoch after which is possible to phase-connect the data.}
Errors in the  center and bottom panels are at $1\sigma$.}
\label{flux_freq_der}
\end{figure}

\section{Timing analysis}
\label{timing}

%We firstly note that, hereafter, our timing analysis is based on the phase-fitting technique.

In order to study the evolution of the spin frequency from outburst to quiescence (i.e. covering the whole data set) we initially measured the spin frequency in each individual pointing by applying a phase-fitting technique in every observation. The phase of a pulse is defined as $\phi=\phi_0 + \int \nu dt$, where $\nu$ is the spin frequency. 
If the coherence of the signal is maintained between subsequent observations, the data can be be fitted by the polynomial:

\begin{equation}
\label{phase-fit}
\phi (t) = \phi_0 + \nu_0  (t-T_0) + \cfrac{1}{2} \dot{\nu}  (t-T_0)^2 + \cfrac{1}{6}  \ddot{\nu}  (t-T_0)^3 + ... 
\end{equation}

\noindent
where $T_0$ is the reference epoch, $\nu_0$ is the frequency at $T_0$, $\dot{\nu}$ is the spin frequency derivative and $\ddot{\nu}$ is second-order spin frequency derivative  \citep[e.g.][for more details]{dallosso03}.

%We note that here the phase connection is not maintained between consecutive pointings. 
Thanks to the large counting statistics of each single observation, it was possible to obtain accurate measurements of the frequencies by applying the phase-fitting technique to  a number of short time intervals (durations from 300 s to 5 ks, depending on the counting statistics) within each observation and we were able to align the pulse-phases by use only the linear term of Eq.~\ref{phase-fit}.
The frequencies derived in this way are plotted as a function of time in the middle panel of Figure~\ref{flux_freq_der}, 
while in the top panel we show the flux evolution of \src .

To derive the fluxes plotted in   Figure~\ref{flux_freq_der}, we fitted 
the time-averaged spectra of each  observation with a model consisting of two to three blackbodies (see e.g. \citealt{bernardini09,alford16} for more details).
The interstellar absorption was kept fixed to the value of $5.7\times10^{21}$ cm$^{-2}$, derived from the spectrum of the first {\it {\it XMM-Newton}} observation.  The temperatures that we found for the three blackbodies ($\sim$0.1, 0.3 and 0.5 keV) are 
consistent with those reported in   
\citet{bernardini09} and \citet{alford16}, to which we refer for more details.
The maximum flux observed by {\it {\it XMM-Newton}} during the outburst was $(3.18\pm0.04)\times10^{-11}$ erg cm$^{-2}$ s$^{-1}$ (absorbed flux in the 0.3-10 keV energy range). The flux decreased until about MJD 54500, after which it remained rather constant (see also \citealt{gh07,bernardini11,alford16}).
We found that the flux slowly decreased, finally reaching a constant value of  $(7.5\pm0.2)\times10^{-13}$ erg cm$^{-2}$ s$^{-1}$, which we derived by fitting with a constant the fluxes of all the observations after MJD 54500 (see Fig.~\ref{flux_freq_der}-{\it top panel}). This value is within the range of fluxes measured by {\it ROSAT}, {\it ASCA} and {\it Einstein} before the onset of the outburst ($(5-10) \times 10^{-13}$ erg cm$^{-2}$ s$^{-1}$; \citealt{gotthelf04}).

It is clear from Fig.~\ref{flux_freq_der} that the source timing properties tracked remarkably well the evolution of the  flux. The average spin-down rate was larger during the first 3-4 years, during the outburst decay, and then it decreased while the source was in (or close to) quiescence. 
We can  distinguish two  time intervals, separated at MJD $\sim54000$, in which  a linear fit can approximately describe the  frequency evolution. The slopes of the two linear functions are $(-3.9\pm0.2)\times10^{-13}$  Hz  s$^{-1}$ ($\chi^2_{\nu}/dof=6.7/9$) and $(-1.00\pm0.05)\times10^{-13}$ Hz  s$^{-1}$ ($\chi^2_{\nu}/dof=1.8/24$) before and after MJD 54000, respectively. 
These values represent the long-term averaged spin-down rates,  but the residuals of the linear fits 
indicate that the time evolution of the frequency derivative is more complex. 
To better investigate this behavior, we performed several 
linear fits to small groups of consecutive  frequency measurements. 
We   adopted a moving-window approach by using partially overlapping sets of points.
In this way we obtained the $\dot \nu$  values plotted in the bottom panel of Figure~\ref{flux_freq_der}. They show a highly variable spin-down  rate, especially during the outburst decay, when it ranged  from $-4.5\times10^{-13}$ Hz  s$^{-1}$ to $-0.5\times10^{-13}$ Hz  s$^{-1}$.

\begin{figure*}
\center
\subfigure{\includegraphics[width=12.0cm]{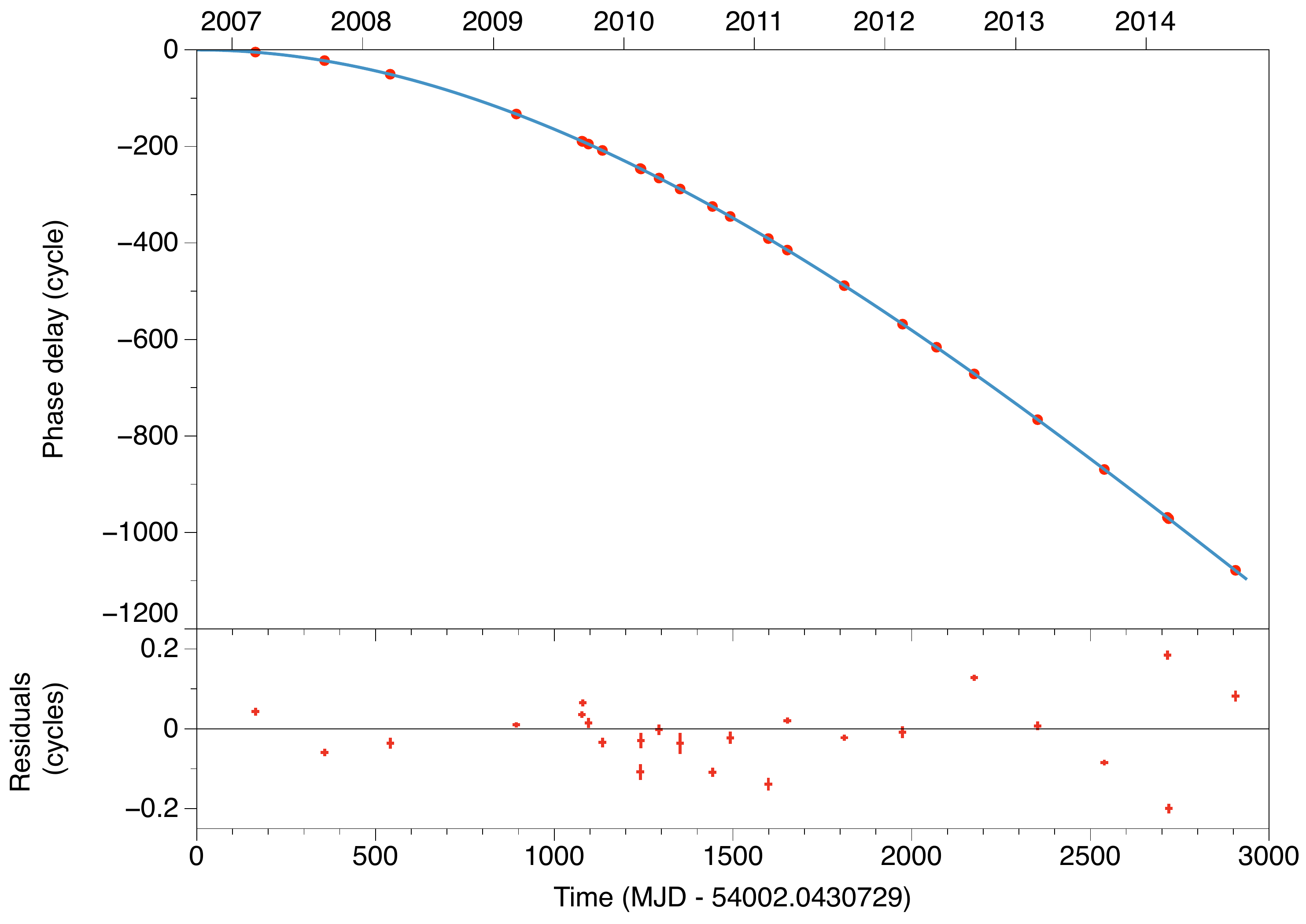}}
\caption{Phase-connection of $\sim3000$ days of {\it {\it XMM-Newton}} and {\it {\it Chandra}} data (observations from 13 to 38) using a third order polynomial function. {\it Top:} the red points are the measured phases, one for each observation, and the solid line is the best fit model;  
{\it bottom:} residuals with respect to the best fit model.}
\label{best-fit}
\end{figure*}

Phase-coherent timing solutions for \src\ have been reported for the initial part of the outburst  \citep{ibrahim04,camilo07}. 
We tried to phase-connect all the {\it XMM-Newton} and {\it Chandra} observations, but this turned out to be rather difficult due to the large timing noise. 
However, we were able to find a   phase-connected  solution for the data during quiescence (i.e. all the observations obtained after MJD 54100),  as follows.
For each observation, we folded the EPIC (pn plus MOS) or {\it {\it Chandra}} data at a  frequency of 0.18048 Hz (corresponding to $P=5.54078$ s,  the average spin period after MJD 54100). For each observation the phase of the pulsation was then derived by fitting a constant plus a sinusoid to the folded pulse profile in the 0.3-10 keV energy range. 
We initially aligned, with only the  linear term in Eq.~\ref{phase-fit},
the {\it {\it XMM-Newton}} observations 18 and 19 that were the most closely spaced ($\sim$2 days).
Then, we included one by one the other observations, as the uncertainty on the best-fit parameters became increasingly smaller allowing us to connect more distant points. We included higher order derivatives only if the improvement in the fit was significant in the timing solution. 
After the inclusion of {\it Chandra} observations 21 and 22, 
the quadratic term became statistically significant,
while the third order polynomial term was needed 
after the inclusion of observations 25 and 26.
The best fit parameters of the final solution are reported in Table~\ref{timing_sol} 
and the fit is shown in Figure~\ref{best-fit}. 
The fit with $\nu$, $\dot{\nu}$ and $\ddot{\nu}$ has $\chi_\nu^2$= 65.7 (for 20 dof). Such a large  value reflects the presence of a strong timing noise.
In fact, the residuals  shown in the lower panel of Figure~\ref{best-fit} indicate significant deviations from the best fit solution, especially during the last 1000 days, when they are as large as  $\sim$0.2 cycles in phase.

\begin{figure}
\center
\includegraphics[width=8.3cm]{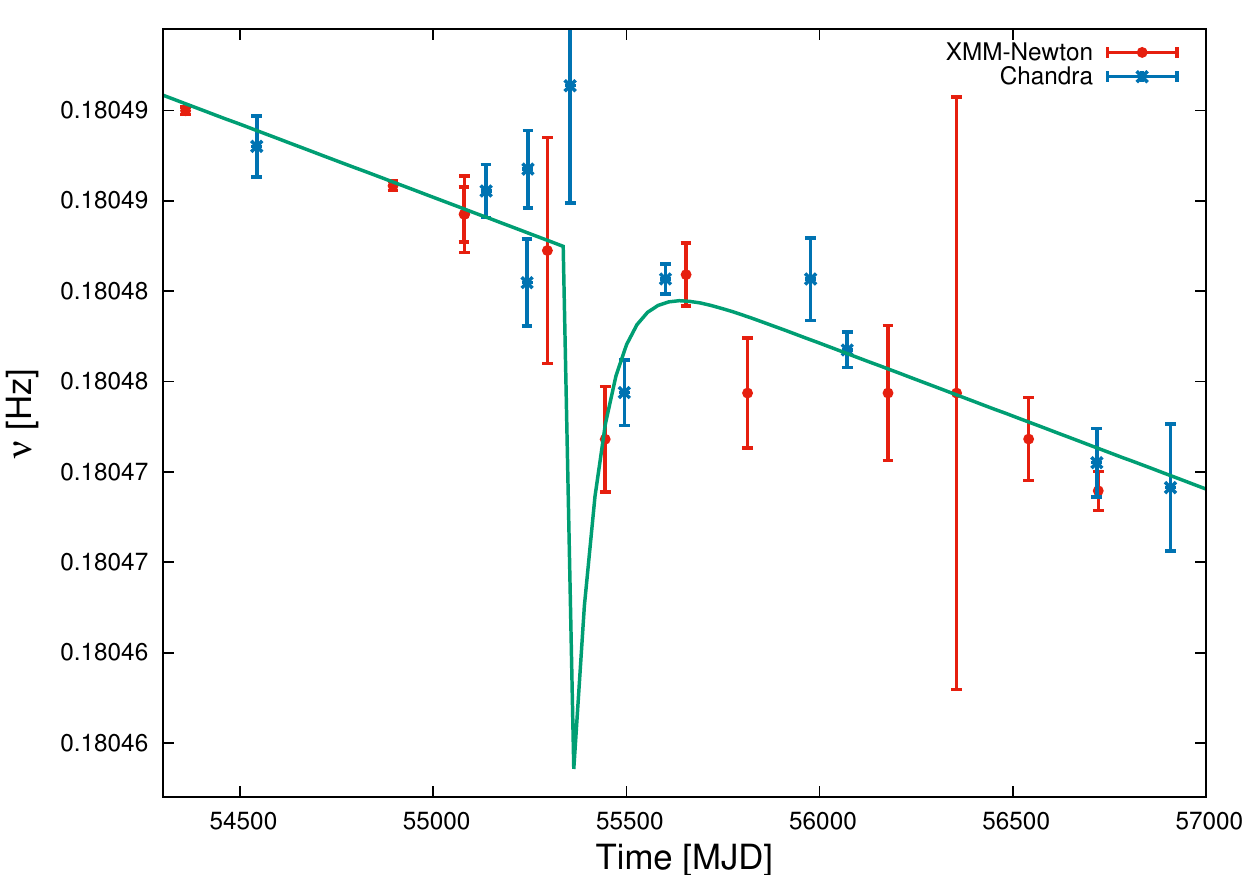}
\caption{Frequencies measured around the time of the possible anti-glitch. The solid line is the best-fit discussed in the text. }
\label{glitch_fit}
\end{figure}

Some timing irregularity occurred  also when the source was in quiescence.  In particular, around MJD $\sim$55400 the spin-down rate was much larger than the quiescent average value and larger than that seen during the outburst decay. Quite remarkably, also a spin-up episode was detected  (see Figure~\ref{flux_freq_der}-\textit{bottom}). 
This is better illustrated in Figure~\ref{glitch_fit} which shows the frequency measurements around this time. 
Unfortunately, the sparse coverage and the large error bars of some points do not allow us to establish whether this was a sudden event,  like an anti-glitch, or simply due to an increased timing noise episode.
Assuming that the time irregularity is an anti-glitch, we fitted the data in the time range MJD 54300--57000, with the following simple model: 

\smallskip
$\nu(t) = \nu_0 + \dot{\nu_0}\cdot t $    ~~~~~~~~~~~~~~~~~~~~~~~~~~~~~~~~~~~~~for  $t<t_g$

\smallskip
$\nu(t) = \nu_0 + \dot{\nu_0}\cdot t + \Delta\nu \cdot e^{(-(t-t_g)/\tau)}$   ~~~ for  $t>t_g$

\smallskip
\noindent
where  $\tau$ is the decay time and $t_g$ is the time of the glitch, which we kept fixed in the fit. If the glitch occurred immediately after observation 25 ($t_g=55354$), we obtained a good fit ($\chi^2_{\nu} =  1.14$ for 21 dof, shown by the solid line in Figure~\ref{glitch_fit}) with $\Delta\nu = (6.5\pm5.8)\times10^{-5}$ Hz, $\tau=51\pm21$ days, $\nu_0 = 0.18093(3) $ Hz and  $\dot{\nu_0} =-9.4 (3)\times 10^{-14}$ Hz s$^{-1}$.
If instead the glitch occurred at observation 26 ($t_g=55444$), we obtain $\Delta\nu < 1 \times10^{-4}$ Hz and $\tau < 200$ days ($3\sigma$ upper limits).

\section{Discussion}

Variations in the spin-down rate are not uncommon in magnetars and have been observed both in transient and persistent sources.
They are believed to originate from changes in the magnetosphere geometry and particles outflow which produce a varying torque on the neutron star. Since also the emission properties from magnetars depend on the evolution of their dynamic magnetospheres, some correlation between spin-period evolution and radiative properties is not surprising.

\begin{table}
  \begin{center}
\footnotesize
   \caption{Best-fit timing solution of the {\it {\it XMM-Newton}} and {\it {\it Chandra}} observations. Errors   are at 1$\sigma$.} 
      \label{timing_sol}
   \begin{tabular}{l c c }
\hline 
 Parameter & & Units \\
\hline
Time range &  54165--56908 & MJD                         \\
T$_{0}$$^a$ &  54002.0430729 & MJD                         \\
$\nu_0$ & 0.1804821(1) & Hz   \\
$\dot{\nu}$ & $-4.9(2)\times10^{-14}$ & Hz s$^{-1}$  \\
$\ddot{\nu}$ & $1.8(1)\times10^{-22}$ & Hz s$^{-2}$ \\
$P$ & 5.540716(3) & s  \\
$\dot{P}$ & $1.51(7)\times10^{-12}$ & s s$^{-1}$  \\
$\ddot{P}$ & $-5.5(4)\times10^{-21}$ & s s$^{-2}$ \\

$\chi_\nu^2 (dof)$ &65.7 (20)& \\
  \hline
\end{tabular}
\end{center}
$^a$ Reference epoch.
\end{table}

The most striking examples, among persistent magnetars, are given by SGR 1806--20 and 1E 1048.1--5937.
The average spin-down rate of SGR 1806--20, as well as its spectral hardness, increased in the $\sim 4$ years of
enhanced bursting activity which led to the giant flare of December 2004 \citep{mereghetti05}. 
However, a further increase (by a factor of 2--3) of the long term spin-down rate occurred both in 2006 and 2008, while the flux and bursting rate showed no remarkable changes \citep{younes15}.
In 1E 1048.1--5937, significant enhancements of the spin-down rate,
which  then subsided through repeated oscillations, have been observed to lag the occurrence of    X-ray outbursts
\citep{archibald15}.  
Other persistent magnetars, for which phase-coherent timing solutions extending over several years could be
mantained,  showed   $\dot \nu$ variations and/or glitches, sometimes (but not always) related to changes in the source flux and the emission of bursts \citep[e.g.][]{dib14}.

Transient magnetars offer, in principle, the best opportunity to investigate the correlations between the variations in the spin-down rate and the radiative properties. However, the observations of transient magnetars carried out up to now have shown a variety of  different behaviors. Furthermore, for many of them, no detailed information is available on the spin-down  during the quiescent state, that instead in this work we now have found for \src\ . 
No firm conclusion on the evolution of the spin-down rate could be derived from the two outbursts of
CXOU J164710.2--455216,
for which a positive  $\ddot\nu$ was reported only during the decay of the first outburst, while the insufficient time coverage 
prevented such a measure for the second one  \citep[][]{rodriguez14}. 
A positive $\ddot\nu$ was reported for  both Swift J1822.3--1606 (which went in outburst in July 2011 and was subsequently monitored for about 500 days; \citealt{rodriguez16}), as well as for SGR J1935+2154 (outburst in July 2014, time coverage $\sim 260$ days; \citealt{israel16}), and, tentatively,
also for SGR 0501+4516 (for this source observations actually covered part of the quiescent state but phase connection along the entire dataset could not be ensured; \citealt{camero14}). 
On the other hand, an increase of the spin-down rate during the outburst decay was reported for SGR J1745--29 \citep[][]{kaspi14,cotiZelati15}.

Our  analysis of {\it XMM-Newton} and {\it Chandra} data spanning  11 years has shown that in the transient magnetar \src\ the spin frequency evolution  tracked remarkably well the  luminosity state. During  the outburst decay, the average spin-down rate was $(-3.9\pm0.2)\times10^{-13}$  Hz  s$^{-1}$,  but large variations around this value were seen, as already noticed by  several authors \citep[][]{halpern05,camilo07,bernardini09}. During the long   quiescent state after the end of the outburst, the average spin-down rate was a  factor of $\sim4$ smaller.  Although some timing noise was still present, the variations in $\dot \nu$ were smaller in the quiescent state, except for a few months in Summer 2010. The timing irregularities in that period might have been caused by the occurrence of an anti-glitch, similar to that seen in the persistent  magnetar 1E 2259+586 \citep[][]{archibald13}. { We found that the pulse-shape in the 0.3--10 keV energy range was nearly sinusoidal and the pulse fraction decreased during the outburst decay, as already reported by e.g. \citet{perna08}, \citet{albano10} and \citet{bernardini09}. We note that the pulse-shape remained nearly sinusoidal also during quiescence (see also \citealt{bernardini11,alford16}).}

The spectral properties of magnetars are commonly explained in terms of the twisted magnetosphere model \citep[][]{thompson02b}, according to which part of the magnetic helicity is transferred from the internal to the external magnetic field, which acquires a non-vanishing toroidal component (a twist). 
The currents required to support the twisted external field resonantly up-scatter thermal photons emitted by the star surface, leading to the formation of the power-law tails observed up  to  hundreds of keV. Since twisted fields have a weaker dependence on the radial distance with respect to a dipole, the higher magnetic field at the light cylinder radius results in an
enhanced spin-down rate. The increased activity of magnetars is often associated to the development (or an increase) of a twist, which should lead to higher fluxes, local surface temperature increases, 
harder spectra and larger spin-down rates.
However, this holds for globally twisted fields (meaning that the twist affects the entire external field). The transport of helicity from the interior is mediated by the star crust: in order to occur the crust must yield, allowing a displacement of the field lines. 
Crustal displacements are small compared to the star radius, so the twist is most likely localized to a bundle of field lines anchored on the displaced platelet \citep[][]{beloborodov09}. Once implanted, the twist must necessarily decay to maintain its own supporting currents, unless energy is constantly supplied from the star interior.
The sudden appearance of a localized twist and its subsequent decay can explain some of the observed properties of transient magnetars \citep[][]{beloborodov09,albano10},  including the fact that transient spectra are often thermal, as in the case of \src , since resonant Compton scattering may be not very effective, although the mechanism responsible for the heating of the
star surface is still unclear (either Ohmic dissipation by backflowing currents or deep crustal heating; \citealt{beloborodov09,pons12}). 
If strong enough, a localized twist can still influence the spin-down rate, which is expected to increase first and then decrease as the magnetosphere untwists, as we observed in \src .

\section{Conclusions}

\src\ was the first transient magnetar to be discovered and it is probably one of the best studied. In particular, it has been possible to trace in great detail its spectral properties over the long ($\sim 3$ years) outburst decay and to monitor it during quiescence for several years afterwards. By   investigating  the evolution of its spin frequency with all the available {\it XMM-Newton} and {\it Chandra} data,  we found evidence for two distinct regimes:
during the outburst decay, $\dot{\nu}$ was highly variable in the range $-(2-4.5)\times10^{-13}$ Hz s$^{-1}$, while during  quiescence the   spin-down rate was more stable and had an  average value smaller by a factor $\sim4$. 

This evolution of the spin-down rate  is in agreement with the suggestion that the  outburst of transient magnetars  may be caused by a strong twist of a localized bundle of magnetic field lines \citep{beloborodov09}.  
Evidences for an evolution of $\dot\nu$ in other transient magnetars are far less conclusive, possibly reflecting the fact that,  if the twist is not very strong,  or the twisted bundle too localized, its effect on the spin-down rate are smaller. 
{ A detailed calculation of the spin-down torque for a spatially-limited twisted field requires a full non-linear approach and has not been presented yet. \citet{beloborodov09} discussed a simple estimate, valid for small twists ($\psi<1\ \mathrm{rad}$)
\begin{equation} 
\label{deltamu}
\Delta\mu/\mu\sim (\psi^2/4\pi)\log(u_*/u_\mathrm{LC})\,,
\end{equation} 
where $\Delta\mu$ is the ``equivalent'' increase in the dipole moment produced by the twist and $u$ is the area of the j-bundle, evaluated at the star surface and at the light cylinder. Since $\Delta\dot\nu/\nu \sim 2\Delta\mu/\mu$, a fractional variation of $\dot\nu$ of a factor of $\sim 4$, as observed (see Figure~\ref{flux_freq_der}, lower panel), can not be achieved with a small twist, $\psi<1$. This indicates that the (maximal) twist in \src\ was most probably larger, $\psi\ga 1 \ \mathrm{rad}$, so that equation (\ref{deltamu}) does not hold anymore. A quite large value of the twist in the outburst of \src\ was also inferred by \citet{beloborodov09}, on the (qualitative) basis that only a strong twist can produce a change of the spin-down rate.   }

\addcontentsline{toc}{section}{Bibliography}
\bibliographystyle{mn2e}
\bibliography{biblio}

%\newpage
%\clearpage
%\newpage

%\end{document}

%\begin{document}

\section*{ERRATUM: ``The variable spin-down rate of the transient magnetar XTE\,J1810--197''}

%\maketitle

Prompted by the recent paper by Camilo et al. (2016), we re-examined our phase-connected timing solution for  XTE J1810--197 (Pintore et al. 2016), and we found a flaw in the procedure to compute the errors during some steps of our analysis. 
Due to  this mistake, the phase-connected solution on 3000 days of X-ray data (reported in Tab. 2 and  Fig. 2 of Pintore et al. 2016) is wrong.

With the new analysis of the data, we can phase-connect 13 observations with a good fit ($\chi^2_{\nu}$ (dof)$=0.9 (9)$; solution 1 in Tab.3 and Fig.4-top), from MJD 55079 to MJD 55814 (observations from 18 to 30 of Pintore et al. 2016). 
The inclusion of  also the two observations at  MJD 55976 and  MJD 56071 (observations 31 and 32) yields  best fit parameters (solution 2 in Tab.3 and Fig.4-bottom) consistent with those obtained by  Camilo et al. (2016) for the same set of observations, but with a higher $\chi_\nu^2$ with respect to solution 1.  

The table and figures reported here supersede Tab. 2 and Fig. 2 of Pintore et al. (2016). We note that these changes do not affect the main conclusions of that paper.

\begin{figure}
\center
\subfigure{\includegraphics[width=8.0cm]{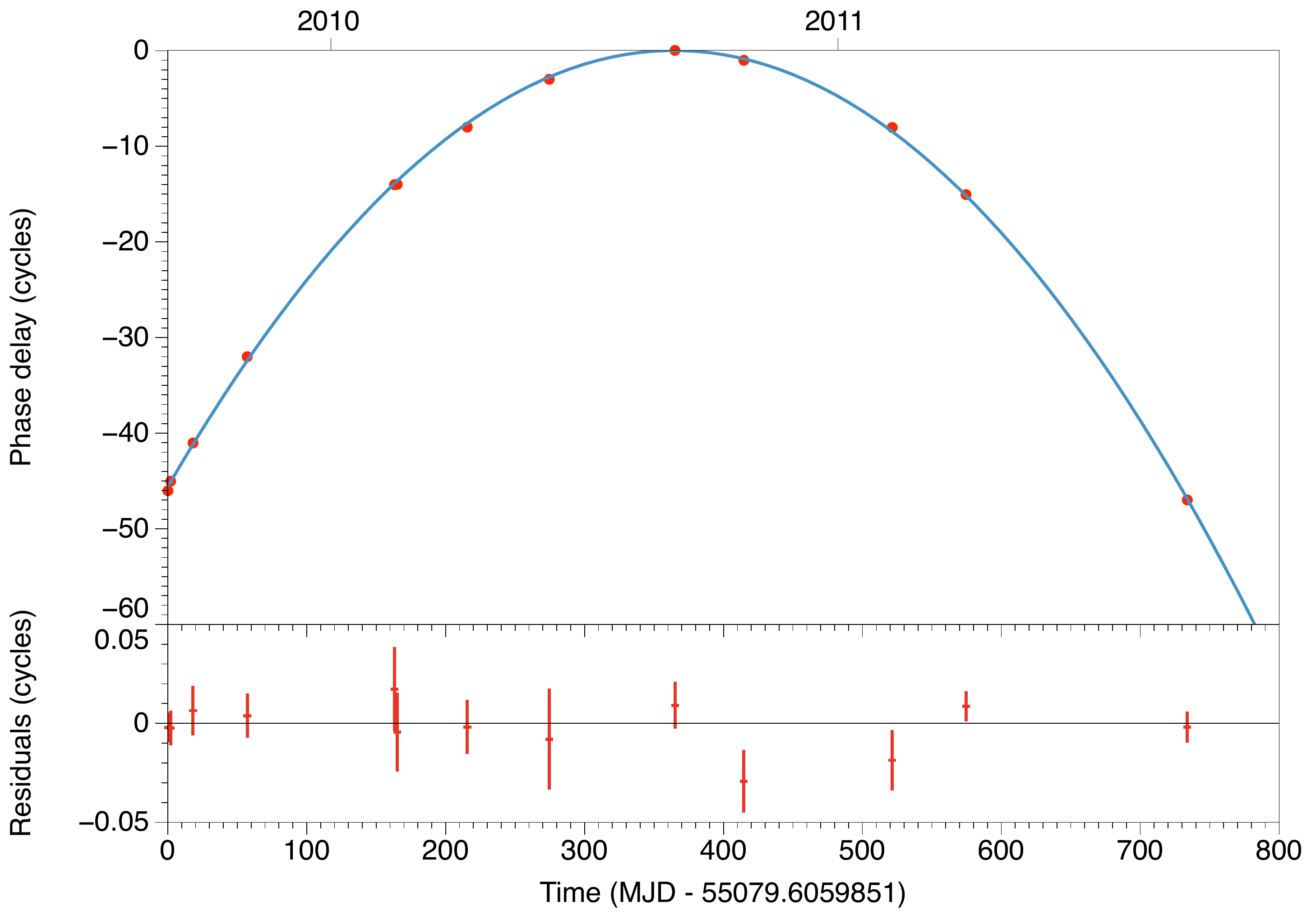}}
\subfigure{\includegraphics[width=8.0cm]{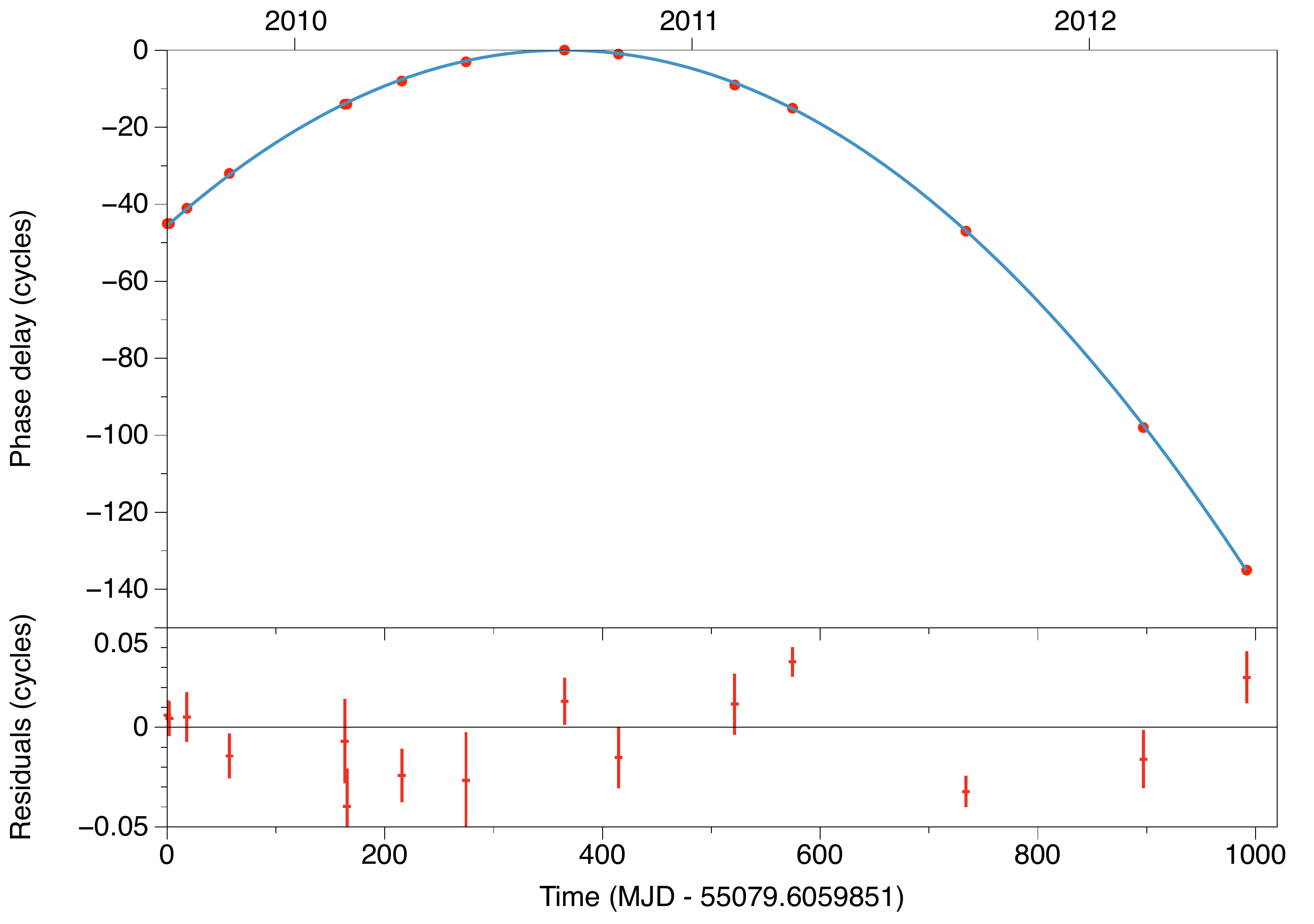}}
\caption{Phase-connection of $\sim800$ (left) and $\sim1000$ (right) days of {\it {\it XMM-Newton}} and {\it {\it Chandra}} data using a third order polynomial function.  {\it Top panels:} the red points are the measured phases, one for each observation, and the solid line is the best fit model;  
{\it bottom panels:} residuals with respect to the best-fit model.}
\label{best-fit}
\end{figure}

\begin{table}
\begin{center}
\caption{Best-fitting timing solutions for 13 (solution 1) and 15 (solution 2) {\it XMM-Newton} and {\it Chandra} observations. Errors are at 1$\sigma$.}
\scalebox{0.89}{\begin{minipage}{24cm}
        \footnotesize
   \begin{tabular}{l c c }
            \hline 
 Parameter &  Solution 1 & Solution 2   \\
\hline
Time range (MJD) &   55079--55814& 55079--56071 MJD      TDB                         \\
T$_{0}$$^a$ (MJD      TDB) &  55444.0& 55444.0                    \\
$\nu_0$ (Hz) & 0.18048121335(44)& 0.18048121599(27)    \\
$\dot{\nu}$ (Hz s$^{-1}$)& $-9.2059(16)\times10^{-14}$ &  $-9.2085(16)\times10^{-14}$  \\
$\ddot{\nu}$ (Hz s$^{-2}$) &  $5.7(3)\times10^{-23}$& $3.80(13)\times10^{-23}$ \\
$P$ (s) & 5.540742892(14)&  5.540742811(8)  \\
$\dot{P}$ (s s$^{-1}$) & $2.8262(5)\times10^{-12}$ & $2.8270(5)\times10^{-12}$  \\
$\ddot{P}$ (s s$^{-2}$) & $-1.75(9)\times10^{-21}$&  $-1.16(4)\times10^{-21}$ \\
$\chi_\nu^2$ (dof) & 0.9 (9) &5.0 (11) \\

  \hline
        \end{tabular}
    \end{minipage}}
        \end{center}     
 $^a$ Reference epoch.\\
\end{table}

\section*{References}
Camilo F., Ransom S. M., Halpern J. P., Alford J. A. J., Cognard I., Reynolds J. E., Johnston S., Sarkissian J., van Straten W., 2016, ApJ, 820, 110

\noindent Pintore F., Bernardini F., Mereghetti S., Esposito P., Turolla R., Rea N., Coti Zelati F., Israel G. L., Tiengo A., Zane S., 2016, MNRAS, 458, 2088

\end{document}